\documentclass[article,nofootinbib,preprintnumbers,tightenlines,twocolumn,superscriptaddress]{revtex4}
\usepackage{graphicx,amsfonts,color,comment,amsmath,hyperref,float}
\usepackage{amssymb}	

\usepackage{mathrsfs,amssymb}  
\usepackage{cancel}
\usepackage[normalem]{ulem}

\newcommand{\be}{\begin{equation}}
\newcommand{\ee}{\end{equation}}
\newcommand{\bea}{\begin{eqnarray}}
\newcommand{\eea}{\end{eqnarray}}

\newcommand{\com}[1]{\textcolor{red}{#1}}

\usepackage{enumitem}
\setlist[itemize]{leftmargin=3mm}

\begin{document}

\title{Reflections On the Anomalous ANITA Events: \\The Antarctic Subsurface as a Possible Explanation}

\author{Ian M. Shoemaker}
\affiliation{Center for Neutrino Physics, Department of Physics, Virginia Tech, Blacksburg, VA 24061, USA}

\author{Alexander Kusenko}
\affiliation{Department of Physics and Astronomy, University of California, Los Angeles\\
Los Angeles, CA 90095-1547, USA}
\affiliation{Kavli Institute for the Physics and Mathematics of the Universe (WPI), UTIAS\\
The University of Tokyo, Kashiwa, Chiba 277-8583, Japan}

\author{Peter Kuipers Munneke}
\affiliation{Institute for Marine and Atmospheric Research,
Utrecht University, Utrecht, The Netherlands}

\author{Andrew Romero-Wolf}
\affiliation{Jet Propulsion Laboratory, California Institute of Technology, Pasadena, CA 91109 USA}

\author{Dustin M. Schroeder}
\affiliation{Departments of Geophysics and Electrical Engineering, Stanford University, Stanford, CA, USA}

\author{Martin J. Siegert}
\affiliation{Grantham Institute and Department of Earth Science and Engineering, Imperial College London, South Kensington, London SW7 2AZ, UK}

\date{\today}
\begin{abstract}
The ANITA balloon experiment was designed to detect radio signals initiated by neutrinos and cosmic ray air showers. These signals are typically discriminated by the polarization and phase inversions of the radio signal. The reflected signal from cosmic rays suffer phase inversion compared to a direct tau neutrino event. In this paper we study sub-surface reflection, which can occur without phase inversion, in the context of the two anomalous up-going events reported by ANITA. We find that subsurface layers and firn density inversions may plausibly account for the events, while ice fabric layers and wind ablation crusts could also play a role. This hypothesis can be tested with radar surveying of the Antarctic region in the vicinity of the anomalous ANITA events. Future experiments should not use phase inversion as a sole criterion to discriminate between downgoing and upgoing events, unless the subsurface reflection properties are well understood. 
\end{abstract}


\preprint{IPMU19-0070}
\maketitle

%

\section{Introduction}

The Antarctic Impulsive Transient Antenna (ANITA) collaboration reports two unusual steeply pointed up-going air showers with energies near the EeV ($10^{18}$ eV) scale~\cite{Gorham:2016zah,Gorham:2018ydl}.  Additionally ANITA has observed $\sim 30$ events from cosmic rays (CRs)~\cite{Gorham:2016zah,Gorham:2018ydl,2010PhRvL.105o1101H}. Most of the CR events appearing to originate from the Earth display a characteristic phase reversal 
consistent with the interpretation that the signal originated from a downward-going CR-initiated extensive air shower (EAS) reflected by the Antarctic surface.  However, the two anomalous upgoing EASs reported by ANITA~\cite{Gorham:2016zah,Gorham:2018ydl} lack phase inversion, and they appear to be inconsistent with such surface reflections.  

Qualitatively, these events look like air showers initiated by an energetic particle that emerges from the ice. It has been proposed that tau neutrinos interacting in the ice could produce a tau lepton that exits upward to the atmosphere and subsequently decays producing an air shower.  Since radio signal from such an event undergoes no reflection, it would be consistent with the observed lack the phase inversion. The problem with this hypothesis, however, is that neutrinos with EeV energies are expected to interact inside Earth with a high probability. For the angles inferred from the observed events, the ice would be well screened from upgoing neutrinos by the underlying layers of Earth.  While the neutrino interaction cross sections at high energies are uncertain~\cite{Kusenko:2001gj}, the observed events would require neutrino fluxes well in excess of upper limits from Pierre Auger Observatory and IceCube~\cite{2018arXiv181107261R}.
A number of new physics explanations for the anomaly have been proposed~\cite{Cherry:2018rxj,Anchordoqui:2018ucj,Huang:2018als,Dudas:2018npp,Connolly:2018ewv,Fox:2018syq,Collins:2018jpg,Chauhan:2018lnq,Anchordoqui:2018ssd,Heurtier:2019git,Hooper:2019ytr,Cline:2019snp}.

In this {\em Letter}, we study the possibility that the mysterious events are explained by the radio signals originating from downward-going CR-initiated EAS reflected by some subsurface features in the Antarctic ice which allow for reflections without a phase inversion. Phase inversion occurs when the radio waves traveling in a medium with a low index of refraction (air) reflect from an interface with a medium that has a high index of refraction (ice).  However, if the reflection occurs below the surface, for example from an interface of a high-density layer on top and a low-density layer on the bottom, there is no phase inversion.  We will identify the  properties of the Antarctic ice that are required for the radio signal from an ordinary CR air shower to undergo a reflection without a phase inversion, and we will also identify the features known to exist in Antarctic ice that can be responsible for such reflections capable to explain the ANITA events. 



\begin{figure}[t!]
\includegraphics[angle=0,width=.45\textwidth]{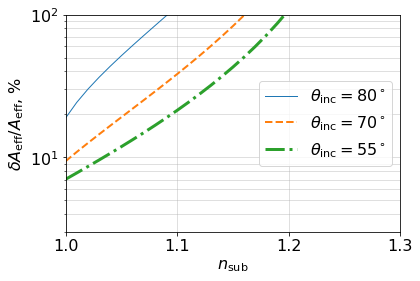}
\caption{The required area coverage of a subsurface reflector as a function of the sub-surface index of refraction, $n_{{\rm sub}}$. Three incidence angles are shown: incident angles of 55$^{\circ}$ (roughly corresponding to the anomalous event~\cite{Gorham:2018ydl}), 70$^{\circ}$ (roughly the average angle for ANITA CRs), and 80$^{\circ}$ (which is on the high end of incident angles for the ANITA CR events). Here it is assumed that the surface has $n=1.3$.}
\label{fig:sub}
\end{figure}

{\it General Features of Subsurface Reflectors.}
ANITA reports 33 events with phases consistent with the expectations from cosmic ray induced ultra-high energy cosmic ray (UHECR) EAS events~\cite{Gorham:2016zah,Gorham:2018ydl,2010PhRvL.105o1101H,Allison:2018cxu}. We compute the number of events above detection threshold $E_{{\rm thr}}$ in an observing time, $T$, reflecting from either surface of subsurface features with area coverage, $A_{{\rm eff}}$, as 
\bea
N&\simeq& A_{{\rm eff}}T\times \int_{E_{{\rm thr}}}^{\infty}~ \Phi(E) dE \\ \nonumber
&=& A_{{\rm eff}}T \times \frac{\Phi_{0}}{(\gamma-1)E_{0}}~\left(\frac{E_{{\rm thr}}}{E_{0}}\right)^{1-\gamma},
\eea
where we take the CR flux to be a power-law, $\Phi(E) \simeq \Phi_{0} (E/E_{0})^{-\gamma}$ with $\gamma \simeq 2.7$.  This allows us to estimate the total number of ordinary EAS events from the surface reflection (with phase inversion): 
\be
N_{{\rm CR}} \simeq A_{{\rm eff}} T \times R_{{\rm surf.}} \times \alpha_{{\rm CR}},
\label{eq:1}
\ee
where we define $\alpha_{{\rm CR}}\equiv \frac{\Phi_{0}}{(\gamma-1)E_{0}}~\left(\frac{E_{{\rm thr}}}{E_{0}}\right)^{1-\gamma}$, $R_{{\rm surf.}}$ is the surface reflection coefficient, and $A_{{\rm eff}}$ is the effective area surveyed by ANITA in flight time $T$.

Similarly we estimate the anomalous uninverted radio events from subsurface reflection of EAS.  The subsurface reflections may occur only for incidence angles small enough, so that the power transmitted into the ice at the air-ice interface is significant.  For the firn index of refraction, the power transmitted downward exceeds 80\% if the incidence angle is smaller than $70^\circ$.  
We note that the incident angles of the anomalous ANITA events are well below this upper bound and, in fact, these angles are smaller than the average incidence angle of the cosmic ray events~\cite{2016APh....77...32S}.  {Therefore 
one can estimate the rate of anomalous events as 
\be N_{{\rm anom.}} \simeq \delta A_{{\rm eff}} T \times (1-R_{{\rm surf.}})^{2}~R_{{\rm sub.}} \times  \alpha_{{\rm CR}}
\label{eq:2}
\ee
where $R_{{\rm sub.}}$ is the subsurface reflection coefficient, and $\delta A_{{\rm eff}}$ is the area of the reflecting subsurface.~\footnote{{\it A priori}, there is no reason to assume that $\delta A_{{\rm eff}}$ is small. In general, $\delta A_{{\rm eff}}$ could exceed $ A_{{\rm eff}}$, especially if several layers at different depths are contributing to the subsurface reflections. However, to explain the ANITA events, only a small fraction of ice needs to host the reflecting features.}  The estimates in Eqs.~(\ref{eq:1}) and (\ref{eq:2}) imply that the subsurface features should occupy an area 
\be
\left(\frac{\delta A_{{\rm eff}}}{A_{{\rm eff}}} \right)  \simeq \frac{R_{{\rm surf.}}}{(1-R_{{\rm surf.}})^{2}~R_{{\rm sub.}}}~\left(\frac{N_{{\rm anom.}}}{N_{{\rm CR}}}\right),
\label{eq:abundance}
\ee
where in order to account for ANITA's observations one needs, $N_{{\rm CR}} = 33$ and $N_{{\rm anom.}} =2$. We plot the requisite area estimate from Eq.~(\ref{eq:abundance}) in Fig.~\ref{fig:sub} as a function of the subsurface index of refraction assuming that the top layer has $n= 1.3$. }

In summary, a relevant candidate subsurface feature needs to satisfy the following requirements:

\begin{enumerate}[leftmargin=4mm]

\item {In accordance with the estimate in Fig.~\ref{fig:sub}, $\gtrsim 7\%$ of the area should host a reflector at some depth.}

\item There should not be significant attenuation for the EAS radio pulse above the reflecting feature, since this would render the signal undetectable. {Roughly, if the detected anomalous event was attenuated by $\lesssim 0.2$, the resulting field amplitude would drop below the trigger threshold~\cite{2018arXiv181107261R}}. Since the attenuation length for radio waves in ice is (1.2--1.5) km (with some temperature dependence~\cite{2012E&PSL.359..173M}) for the frequencies probed by ANITA, this requirement is satisfied by any features not obstructed by an overlying layer of liquid water.  (The attenuation length in liquid water is much shorter~\cite{Ray:72}.) 

\item The reflection must occur without a phase inversion. A subsurface interface with a higher index of refraction above and a lower index of refraction below can reflect a signal without a phase inversion.  Likewise, multiple layers of variable index of refraction can reflect a signal without a phase shift~\cite{Tikhonravov:97}.


\item Given the wavelengths ANITA is sensitive to, the subsurface layer above the reflecting interface needs to be sufficiently thick, although the layer below the interface can be quite small ~\cite{siegert_2016}. Similarly, the features should be $>$20 m in diameter~\cite{doi:10.1029/2004JB003222}. Lastly these surfaces likely need to be tilted  with respect to the surface, such that double reflections are rare. 
 
\end{enumerate}


Given these requirements, we now proceed to investigate which glaciological candidates may have the correct distribution and reflective properties. 
 





%
%

\begin{figure*}[t!]
\begin{centering}
\includegraphics[angle=0,width=.45\textwidth]{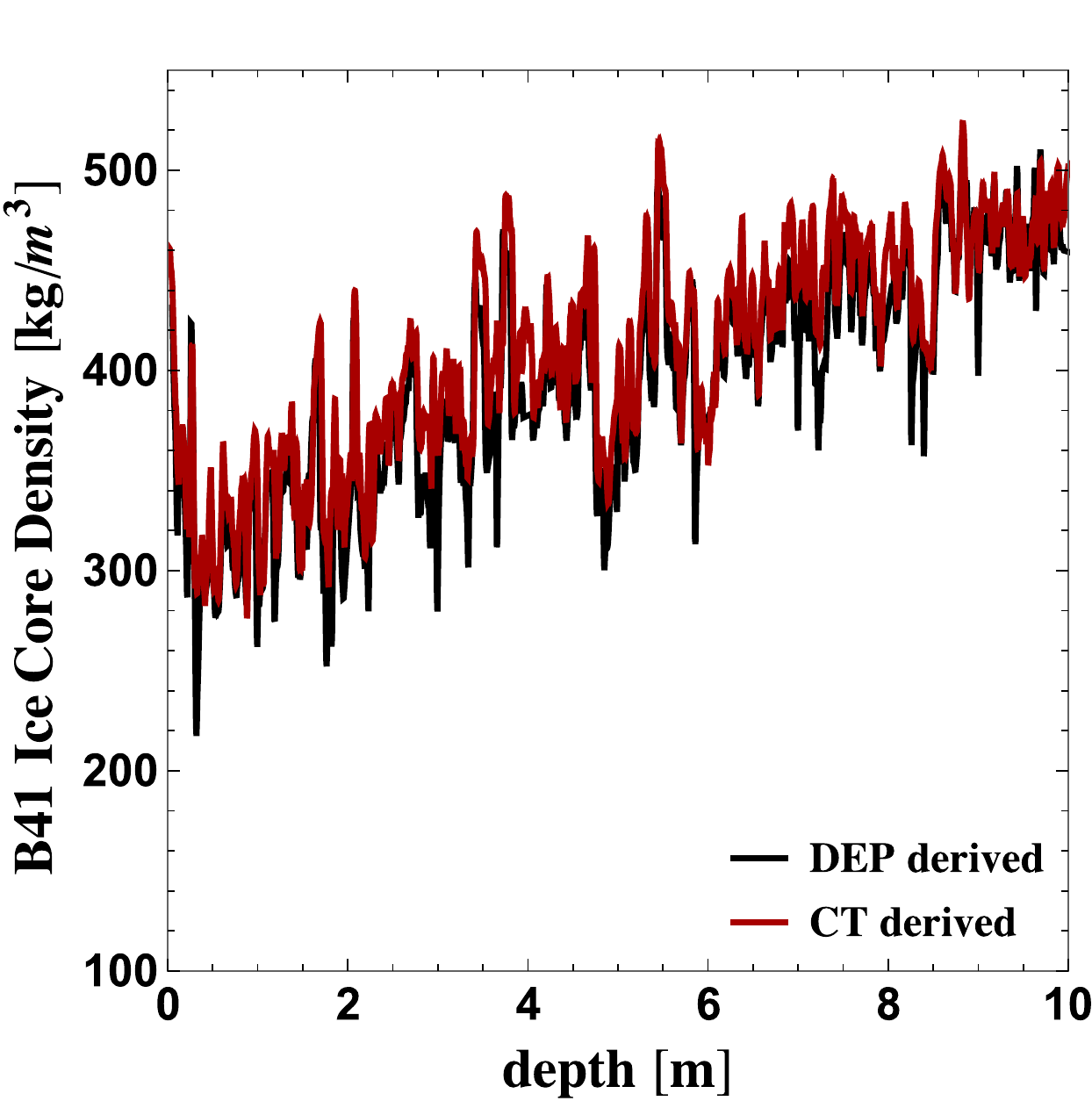}
\includegraphics[angle=0,width=.45\textwidth]{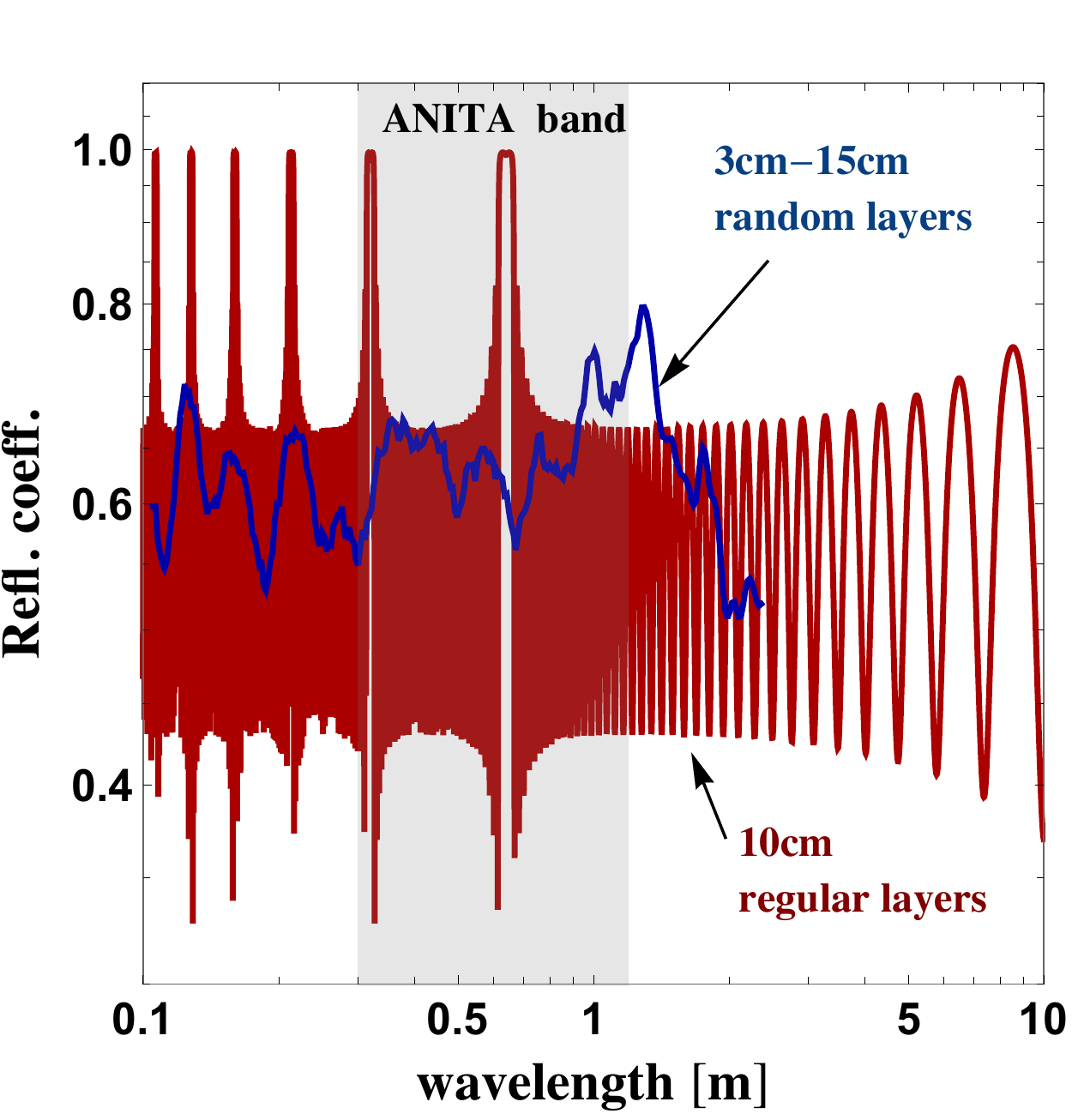}
\end{centering}
\caption{ 
{\it Left panel}: An example Ice Core sample from the East Antarctic Plateau ~\cite{doi:10.1002/2016JF003919}. Here the black curve shows the density from a dielectric profiling (DEP) technique, while the red shows the result from high-resolution X-ray computer tomography (CT).
{\it Right panel}: Reflection coefficients (in power) for scattering from a multilayered medium as a function of wavelength. The red curve is calculated assuming regularly spaced layers of 10 cm thickness, whereas the blue curve assumes layers whose thickness is randomly chosen between 3-15 cm. 
 }

\label{fig:firn}

\end{figure*}


{\it Subsurface features } that may have the right properties to account for the anomalous ANITA events include several possibilities.  

\begin{itemize}

\item[(a)] {\bf Double Layers}: Intriguingly, the work of Ref.~\cite{arcone_spikes_hamilton_2005} finds direct evidence for reflective surfaces without phase inversions. In particular, they find evidence for ``thin double layers of ice over hoar" which have reflection without inversion, and conclude that they are ``extensive" throughout West Antarctica. They model the observed reflections as high-permittivity ice sitting above low-permittivity hoar. The modeling done indicates that hoar thickness fluctuations is a major driver of the phase of the wavelet. These results were obtained with 400 MHz short pulse radar (ANITA is in the 200-800 MHz range).

\item[(b)]{\bf Firn Density Inversions 
}: In Ref.~\cite{Kuipers,Kuipers2}, the authors produced an estimate of snow and firn density (in the top 100 m of depth) on Antarctica for the period 1979-2017 at a horizontal resolution of (27 $\times$ 27) ${\rm km}^{2}$ and a temporal resolution of 10-15 days, using a firn model that includes not only compaction, but also firn hydrology including melt, percolation and refreezing. The model in~\cite{Kuipers,Kuipers2} has been evaluated at the two locations of the observed ANITA events. Although there are  minor variations in density due of dependence of densification on temperature following the annual cycle, these variations are quite small, and are not large enough to explain the ANITA observations. However, there are additional firn features not accounted for in this model. 

For example, ice core samples show substantial density and permittivity variations. We show as an example in the left panel of Fig.~\ref{fig:firn} the density profile from the East Antarctic Plateau ~\cite{doi:10.1002/2016JF003919}. These density variations are quite substantial and may constitute a plausible candidate for the ANITA events, and they are rather common in the area sampled in Ref.~\cite{doi:10.1002/2016JF003919}. 

It is well-known that the constructive interference effects from scattering on a medium consisting of multiple thin layers can produce a large reflection coefficient at some wavelengths~\cite{1977ApOpt..16...89V} (see also \cite{Pirozhkov_2015,doi:10.1063/1.343131} for generalizations).  The ice core sample displayed in Fig.~\ref{fig:firn} has just this structure, consisting of a number of thin layers with small index of refraction differences. Motivated by this, we display two calculations of the reflection coefficients in Fig.~\ref{fig:firn}. In one case (the red curve) we compute reflection from a medium consisting of regularly spaced layers of 10 cm thickness. This results in a sharp resonance feature, as expected~\cite{1977ApOpt..16...89V}.   In contrast, the blue curve assumes layers whose thickness is randomly chosen between 3-15 cm. In both instances the layers alternating refraction indices, chosen between $n_{1} = 1.3$ and $n_{2} = 1.6$ for a $60^{\circ}$ incidence angle.~We have explicitly computed the phase change in reflections from regular and random layers, and found that the phase shifts are close to zero for the range of wavelengths with maximal reflectivity, in agreement with the results of Ref.~\cite{Tikhonravov:97} for regular layers.  We note that reflections from multiple layers will induce a time delay, impacting the measured time profile of the pulse. This makes a multi-layer reflection interpretation of the event reported in Ref.~\cite{Gorham:2018ydl} unlikely, though it may be a candidate explanation for the event in Ref.~\cite{Gorham:2016zah}. 



\item[(c)] {\bf Wind/ablation crusts and Sastrugi
}: These are abundant in megadune regions, and may create low density regions with large grains above higher density snow.  Sastrugi are essentially wind eroded snow, which make irregular ridges on the surface. Given that these regions have a variety of slopes as well, they naturally help explain the lack of double-reflections. By some estimates, as much as 11$\%$ of the East Antarctic Ice Sheet is covered by so-called ``wind glaze''~\cite{2012JGlac..58..633S}, forming a surface with a polished appearance with nearly zero accumulation due to persistent winds.  This could produce both the needed reflection phase and range of angles. {Since these wind crusts are denser than typical snow, they would naturally have larger indices of refraction than typical surface snow.

\item[(d)] {\bf Ice Fabric Layers
}: Ice sheet fabrics are formed as a result of rheology and stress, leading to macroscopic ice crystal alignment.  Some fabrics appear to have the right dielectric properties to produce a reflection without phase inversion {even without index of refraction contrasts~\cite{doi:10.1029/2003JB002425}. In this case it is the contrasts in crystal orientation fabric that source strong reflections~\cite{doi:10.1029/2003JB002425}}. The spatial distribution of ice-sheet fabric is not very well known since ice cores are restricted in number and distribution across Antarctica~\cite{Siegert}. There is some indication that ice fabric layers are more widespread than originally believed~\cite{Siegert,siegert_fujita_2001,SIEGERT2000227}, which make it plausible that the distribution is widespread enough to produce the observed reflections.

\item[(e)] {\bf Subglacial Lakes
}: Most lakes appear to be hydrostatically sealed, and therefore lack an air-water interface which would otherwise provide a useful reflecting surface without phase inversion.  The bottom of the lake could in principle work, but only rather shallow and low conductivity subglacial lake regions~\cite{2015IGRSL..12..443S} would be able to produce a reflection without significant attenuation in water.  Exploiting the time delay and amplitude attenuation in water, radio echo surveys provided the first direct evidence for that subglacial lakes were at least several meters deep~\cite{doi:10.1029/1999JB900271}.  Recent model estimates suggest that $(0.6\pm0.2) \%$ of the Antarctic ice/bed interface is covered by subglacial lakes~\cite{Lakes}. Given that subglacial lakes tend to lack a water-air boundary and that they cover $<1\%$ of the Antarctic area, we do not consider these especially promising candidates for ANITA. {We note however the possibility that impurities in the accreted ice above a lake could in principle produce a higher index of refraction layer above a lower index layer, though is likely uncommon. We note that the ice above Lake Vostok has been found to contain ice fabric contrasts, which could source strong reflections~\cite{MACGREGOR2009222}. If the appearance of ice fabrics is common above subglacial lakes, they may be anomalous reflection candidates. However, such lakes could not be too deep since the signal would be too attenuated otherwise. }

%


%

%
\item[(f)] {\bf Snow covered crevasses/hollow caves/ice bridges}: An ice-to-air boundary would have the correct properties for reflection without phase inversion. However, fumarolic and volcanic ice caves do not seem to be sufficiently common for the ANITA events. Crevasses are common in regions of fast flow, but not are not common in the middle of the ice sheet where the ANITA events are observed. 

 %
 \item[(g)] {\bf Englacial Layers}: At depths beyond the firn (though still $\lesssim 1$ km), dielectric contrasts in the ice may be sufficiently common to explain the events~\cite{doi:10.1029/2004JB003222,doi:10.1029/2005JD006349}. {Moreover, in principle density contrasts in deep englacial layers qualitatively similar to what is displayed in Fig.~\ref{fig:firn} may also source strong reflection coefficients.}
} 
 

\end{itemize}

{\it Future probes} of subsurface reflections can be designed to definitively test the origin of ANITA events and to learn about the properties of Antarctic ice. This can be done (1) using dedicated  survey, and/or (2) digging. Though radio surveying is likely more feasible, we leave a dedicated analysis of these possibilities to future work. 

If subsurface features are ultimately responsible for the ANITA events, the distribution and extent of such features will be important for ANITA going forward. Moreover, a dedicated effort to determine if the ANITA events originate from a subsurface reflector may be relevant for glaciology by providing additional information such as the extent, spatial distribution, and reflective properties of these features.

%
%

{\it To summarize}, we have examined the possibility of the anomalous upward-going ANITA events as originating from ordinary CR initiated air showers. For this to be consistent with the phase information ANITA observes, they must reflect from a subsurface feature without phase inversion. We have surveyed a number of glaciological candidates in order to determine which of these have the correct properties to explain ANITA's observations. We have found that sunsurface double layers and firn density inversions are a plausible explanation of the anomalous events. In order to conclusively test if surface/subsurface glaciological candidates are responsible for the ANITA events, more information is needed on candidate location, fraction of occurrence in the area sampled by ANITA and a more detailed analysis of the ANITA acceptance. We note that, while firn density contrasts appear to be a plausible candidate, one or more of the other glaciological features discussed here may play a sub-dominant role in sourcing strong un-inverted reflections. 

Our results have broad implications for future neutrino and cosmic ray experiments.  Given the possibility of reflections without a phase inversion, future experiments should not use the phase inversion in radio signals as a sole criterion for discriminating between downgoing and upgoing events, unless the properties of the subsurface reflection are well understood.

\section*{Acknowledgements}
We are very grateful to Thomas Laepple for providing the B41 ice core data from Ref.~\cite{doi:10.1002/2016JF003919}. Furthermore, we warmly acknowledge helpful discussions with Kumiko Azuma, Dmitry Chirkin, Francis Halzen, Stefan Ligtenberg, Henning Loewe, and David Saltzberg. 
The work of I.M.S. is supported by the U.S. Department of Energy under the award number DE-SC0019163.
The work of A.K. was supported by the U.S. Department of Energy Grant No. DE-SC0009937, and by the World Premier International Research Center Initiative (WPI), MEXT, Japan. Part of this work was carried out at the Jet Propulsion Laboratory, California Institute of Technology, under a contract with the National Aeronautics and Space Administration. P.K.M. is supported by the Netherlands Earth System Science Centre (NESSC).

\bibliographystyle{JHEP}

\bibliography{nu}

\providecommand{\href}[2]{#2}\begingroup\raggedright\begin{thebibliography}{10}

\bibitem{Gorham:2016zah}
{\scshape ANITA} collaboration, P.~W. Gorham et~al., \emph{{Characteristics of
  Four Upward-pointing Cosmic-ray-like Events Observed with ANITA}},
  \href{https://doi.org/10.1103/PhysRevLett.117.071101}{\emph{Phys. Rev. Lett.}
  {\bfseries 117} (2016) 071101}
  [\href{https://arxiv.org/abs/1603.05218}{{\ttfamily 1603.05218}}].

\bibitem{Gorham:2018ydl}
{\scshape ANITA} collaboration, P.~W. Gorham et~al., \emph{{Observation of an
  Unusual Upward-going Cosmic-ray-like Event in the Third Flight of ANITA}},
  \href{https://arxiv.org/abs/1803.05088}{{\ttfamily 1803.05088}}.

\bibitem{2010PhRvL.105o1101H}
S.~{Hoover}, J.~{Nam}, P.~W. {Gorham}, E.~{Grashorn}, P.~{Allison}, S.~W.
  {Barwick} et~al., \emph{{Observation of Ultrahigh-Energy Cosmic Rays with the
  ANITA Balloon-Borne Radio Interferometer}},
  \href{https://doi.org/10.1103/PhysRevLett.105.151101}{\emph{Physical Review
  Letters} {\bfseries 105} (2010) 151101}
  [\href{https://arxiv.org/abs/1005.0035}{{\ttfamily 1005.0035}}].

\bibitem{Kusenko:2001gj}
A.~Kusenko and T.~J. Weiler, \emph{{Neutrino cross-sections at high-energies
  and the future observations of ultrahigh-energy cosmic rays}},
  \href{https://doi.org/10.1103/PhysRevLett.88.161101}{\emph{Phys. Rev. Lett.}
  {\bfseries 88} (2002) 161101}
  [\href{https://arxiv.org/abs/hep-ph/0106071}{{\ttfamily hep-ph/0106071}}].

\bibitem{2018arXiv181107261R}
A.~{Romero-Wolf}, S.~A. {Wissel}, H.~{Schoorlemmer}, W.~R. {Carvalho}, Jr,
  J.~{Alvarez-Mu{\~n}iz}, E.~{Zas} et~al., \emph{{A comprehensive analysis of
  anomalous ANITA events disfavors a diffuse tau-neutrino flux origin}},
  \href{https://doi.org/10.1103/PhysRevD.99.063011}{\emph{Physical Review D}
  {\bfseries 99} (2019) 063011}
  [\href{https://arxiv.org/abs/1811.07261}{{\ttfamily 1811.07261}}].

\bibitem{Cherry:2018rxj}
J.~F. Cherry and I.~M. Shoemaker, \emph{{Sterile neutrino origin for the upward
  directed cosmic ray showers detected by ANITA}},
  \href{https://doi.org/10.1103/PhysRevD.99.063016}{\emph{Phys. Rev.}
  {\bfseries D99} (2019) 063016}
  [\href{https://arxiv.org/abs/1802.01611}{{\ttfamily 1802.01611}}].

\bibitem{Anchordoqui:2018ucj}
L.~A. Anchordoqui, V.~Barger, J.~G. Learned, D.~Marfatia and T.~J. Weiler,
  \emph{{Upgoing ANITA events as evidence of the CPT symmetric universe}},
  \href{https://doi.org/10.31526/LHEP.1.2018.03}{\emph{LHEP} {\bfseries 1}
  (2018) 13} [\href{https://arxiv.org/abs/1803.11554}{{\ttfamily 1803.11554}}].

\bibitem{Huang:2018als}
G.-y. Huang, \emph{{Sterile neutrinos as a possible explanation for the upward
  air shower events at ANITA}},
  \href{https://doi.org/10.1103/PhysRevD.98.043019}{\emph{Phys. Rev.}
  {\bfseries D98} (2018) 043019}
  [\href{https://arxiv.org/abs/1804.05362}{{\ttfamily 1804.05362}}].

\bibitem{Dudas:2018npp}
E.~Dudas, T.~Gherghetta, K.~Kaneta, Y.~Mambrini and K.~A. Olive,
  \emph{{Gravitino decay in high scale supersymmetry with R -parity
  violation}}, \href{https://doi.org/10.1103/PhysRevD.98.015030}{\emph{Phys.
  Rev.} {\bfseries D98} (2018) 015030}
  [\href{https://arxiv.org/abs/1805.07342}{{\ttfamily 1805.07342}}].

\bibitem{Connolly:2018ewv}
A.~Connolly, P.~Allison and O.~Banerjee, \emph{{On ANITA's sensitivity to
  long-lived, charged massive particles}},
  \href{https://arxiv.org/abs/1807.08892}{{\ttfamily 1807.08892}}.

\bibitem{Fox:2018syq}
D.~B. Fox, S.~Sigurdsson, S.~Shandera, P.~Meszaros, K.~Murase, M.~Mostafa
  et~al., \emph{{The ANITA Anomalous Events as Signatures of a Beyond Standard
  Model Particle, and Supporting Observations from IceCube}},
  \href{https://arxiv.org/abs/1809.09615}{{\ttfamily 1809.09615}}.

\bibitem{Collins:2018jpg}
J.~H. Collins, P.~S. Bhupal~Dev and Y.~Sui, \emph{{R-parity Violating
  Supersymmetric Explanation of the Anomalous Events at ANITA}},
  \href{https://doi.org/10.1103/PhysRevD.99.043009}{\emph{Phys. Rev.}
  {\bfseries D99} (2019) 043009}
  [\href{https://arxiv.org/abs/1810.08479}{{\ttfamily 1810.08479}}].

\bibitem{Chauhan:2018lnq}
B.~Chauhan and S.~Mohanty, \emph{{A common leptoquark solution of flavor and
  ANITA anomalies}},  \href{https://arxiv.org/abs/1812.00919}{{\ttfamily
  1812.00919}}.

\bibitem{Anchordoqui:2018ssd}
L.~A. Anchordoqui and I.~Antoniadis, \emph{{Supersymmetric sphaleron
  configurations as the origin of the perplexing ANITA events}},
  \href{https://doi.org/10.1016/j.physletb.2019.02.003}{\emph{Phys. Lett.}
  {\bfseries B790} (2019) 578}
  [\href{https://arxiv.org/abs/1812.01520}{{\ttfamily 1812.01520}}].

\bibitem{Heurtier:2019git}
L.~Heurtier, Y.~Mambrini and M.~Pierre, \emph{{A Dark Matter Interpretation of
  the ANITA Anomalous Events}},
  \href{https://arxiv.org/abs/1902.04584}{{\ttfamily 1902.04584}}.

\bibitem{Hooper:2019ytr}
D.~Hooper, S.~Wegsman, C.~Deaconu and A.~Vieregg, \emph{{Superheavy Dark Matter
  and ANITA's Anomalous Events}},
  \href{https://arxiv.org/abs/1904.12865}{{\ttfamily 1904.12865}}.

\bibitem{Cline:2019snp}
J.~M. Cline, C.~Gross and W.~Xue, \emph{{Can the ANITA anomalous events be due
  to new physics?}},  \href{https://arxiv.org/abs/1904.13396}{{\ttfamily
  1904.13396}}.

\bibitem{Allison:2018cxu}
{\scshape ANITA} collaboration, P.~Allison et~al., \emph{{Constraints on the
  diffuse high-energy neutrino flux from the third flight of ANITA}},
  \href{https://arxiv.org/abs/1803.02719}{{\ttfamily 1803.02719}}.

\bibitem{2016APh....77...32S}
H.~{Schoorlemmer}, K.~{Belov}, A.~{Romero-Wolf},
  D.~{Garc{\'{\i}}a-Fern{\'a}ndez}, V.~{Bugaev}, S.~A. {Wissel} et~al.,
  \emph{{Energy and flux measurements of ultra-high energy cosmic rays observed
  during the first ANITA flight}},
  \href{https://doi.org/10.1016/j.astropartphys.2016.01.001}{\emph{Astroparticle
  Physics} {\bfseries 77} (2016) 32}
  [\href{https://arxiv.org/abs/1506.05396}{{\ttfamily 1506.05396}}].

\bibitem{2012E&PSL.359..173M}
K.~{Matsuoka}, J.~A. {MacGregor} and F.~{Pattyn}, \emph{{Predicting radar
  attenuation within the Antarctic ice sheet}},
  \href{https://doi.org/10.1016/j.epsl.2012.10.018}{\emph{Earth and Planetary
  Science Letters} {\bfseries 359} (2012) 173}.

\bibitem{Ray:72}
P.~S. Ray, \emph{Broadband complex refractive indices of ice and water},
  \href{https://doi.org/10.1364/AO.11.001836}{\emph{Appl. Opt.} {\bfseries 11}
  (1972) 1836}.

\bibitem{Tikhonravov:97}
A.~V. Tikhonravov, P.~W. Baumeister and K.~V. Popov, \emph{Phase properties of
  multilayers}, \href{https://doi.org/10.1364/AO.36.004382}{\emph{Appl. Opt.}
  {\bfseries 36} (1997) 4382}.

\bibitem{siegert_2016}
M.~G.~P. Cavitte, D.~D. Blankenship, D.~A. Young, D.~M. Schroeder, F.~Parrenin,
  E.~Le{M}eur et~al., \emph{Deep radiostratigraphy of the {E}ast {A}ntarctic
  plateau: connecting the {D}ome {C} and {V}ostok ice core sites},
  \href{https://doi.org/10.1017/jog.2016.11}{\emph{J. Glac.} {\bfseries 62}
  (2016) 323–}.

\bibitem{doi:10.1029/2004JB003222}
M.~E. Peters, D.~D. Blankenship and D.~L. Morse, \emph{Analysis techniques for
  coherent airborne radar sounding: Application to {W}est {A}ntarctic ice
  streams}, \href{https://doi.org/10.1029/2004JB003222}{\emph{Journal of
  Geophysical Research: Solid Earth} {\bfseries 110} }.

\bibitem{doi:10.1002/2016JF003919}
T.~Laepple, M.~H\"orhold, T.~M\"unch, J.~Freitag, A.~Wegner and S.~Kipfstuhl,
  \emph{Layering of surface snow and firn at {K}ohnen {S}tation, {A}ntarctica:
  {N}oise or seasonal signal?},
  \href{https://doi.org/10.1002/2016JF003919}{\emph{Journal of Geophysical
  Research: Earth Surface} {\bfseries 121} 1849}.

\bibitem{arcone_spikes_hamilton_2005}
S.~A. Arcone, V.~B. Spikes and G.~S. Hamilton, \emph{Phase structure of radar
  stratigraphic horizons within {A}ntarctic firn},
  \href{https://doi.org/10.3189/172756405781813267}{\emph{Annals of Glaciology}
  {\bfseries 41} (2005) 10}.

\bibitem{Kuipers}
S.~R.~M. Ligtenberg, M.~M. Helsen and M.~R. van~den Broeke, \emph{An improved
  semi-empirical model for the densification of {A}ntarctic firn},
  \href{https://doi.org/10.5194/tc-5-809-2011}{\emph{The Cryosphere} {\bfseries
  5} (2011) 809}.

\bibitem{Kuipers2}
P.~Kuipers~Munneke, S.~R. Ligtenberg, M.~R. Van Den~Broeke and D.~G. Vaughan,
  \emph{Firn air depletion as a precursor of {A}ntarctic ice-shelf collapse},
  \href{https://doi.org/10.3189/2014JoG13J183}{\emph{Journal of Glaciology}
  {\bfseries 60} (2014) 205–214}.

\bibitem{1977ApOpt..16...89V}
A.~V. {Vinogradov} and B.~Y. {Zeldovich}, \emph{{X-ray and far uv multilayer
  mirrors: Principles and possibilities}},
  \href{https://doi.org/10.1364/AO.16.000089}{\emph{\ao} {\bfseries 16} (1977)
  89}.

\bibitem{Pirozhkov_2015}
A.~S. Pirozhkov and E.~N. Ragozin, \emph{Aperiodic multilayer structures in
  soft x-ray optics},
  \href{https://doi.org/10.3367/ufne.0185.201511e.1203}{\emph{Physics-Uspekhi}
  {\bfseries 58} (2015) 1095}.

\bibitem{doi:10.1063/1.343131}
D.~G. Stearns, \emph{The scattering of x rays from nonideal multilayer
  structures}, \href{https://doi.org/10.1063/1.343131}{\emph{Journal of Applied
  Physics} {\bfseries 65} (1989) 491}
  [\href{https://arxiv.org/abs/https://doi.org/10.1063/1.343131}{{\ttfamily
  https://doi.org/10.1063/1.343131}}].

\bibitem{2012JGlac..58..633S}
T.~A. {Scambos}, M.~{Frezzotti}, T.~{Haran}, J.~{Bohlander}, J.~T.~M.
  {Lenaerts}, M.~R. {van den Broeke} et~al., \emph{{Extent of low-accumulation
  `wind glaze' areas on the East Antarctic plateau: implications for
  continental ice mass balance}},
  \href{https://doi.org/10.3189/2012JoG11J232}{\emph{Journal of Glaciology}
  {\bfseries 58} (2012) 633}.

\bibitem{doi:10.1029/2003JB002425}
K.~Matsuoka, T.~Furukawa, S.~Fujita, H.~Maeno, S.~Uratsuka, R.~Naruse et~al.,
  \emph{Crystal orientation fabrics within the {A}ntarctic ice sheet revealed
  by a multipolarization plane and dual-frequency radar survey},
  \href{https://doi.org/10.1029/2003JB002425}{\emph{Journal of Geophysical
  Research: Solid Earth} {\bfseries 108} }.

\bibitem{Siegert}
B.~{Wang}, B.~{Sun}, C.~{Martin}, F.~{Ferraccioli}, D.~{Steinhage}, X.~{Cui}
  et~al., \emph{{Summit of the East Antarctic Ice Sheet underlain by thick
  ice-crystal fabric layers linked to glacial-interglacial environmental
  change}}, \href{https://doi.org/10.1144/SP461.1}{\emph{Geological Society of
  London Special Publications} {\bfseries 461} (2018) 131}.

\bibitem{siegert_fujita_2001}
M.~J. Siegert and S.~Fujita, \emph{Internal ice-sheet radar layer profiles and
  their relation to reflection mechanisms between {D}ome {C} and the
  {T}ransantarctic {M}ountains},
  \href{https://doi.org/10.3189/172756501781832205}{\emph{Journal of
  Glaciology} {\bfseries 47} (2001) 205–212}.

\bibitem{SIEGERT2000227}
M.~J. Siegert and R.~Kwok, \emph{Ice-sheet radar layering and the development
  of preferred crystal orientation fabrics between {L}ake {V}ostok and {R}idge
  {B}, central {E}ast {A}ntarctica},
  \href{https://doi.org/https://doi.org/10.1016/S0012-821X(00)00121-7}{\emph{Earth
  and Planetary Science Letters} {\bfseries 179} (2000) 227 }.

\bibitem{2015IGRSL..12..443S}
D.~M. {Schroeder}, D.~D. {Blankenship}, R.~K. {Raney} and C.~{Grima},
  \emph{{Estimating subglacial water geometry using radar bed echo specularity:
  Application to {T}hwaites {G}lacier, {W}est {A}ntarctica}},
  \href{https://doi.org/10.1109/LGRS.2014.2337878}{\emph{IEEE Geoscience and
  Remote Sensing Letters} {\bfseries 12} (2015) 443}.

\bibitem{doi:10.1029/1999JB900271}
M.~R. Gorman and M.~J. Siegert, \emph{Penetration of {A}ntarctic subglacial
  lakes by {V}{H}{F} electromagnetic pulses: Information on the depth and
  electrical conductivity of basal water bodies},
  \href{https://doi.org/10.1029/1999JB900271}{\emph{Journal of Geophysical
  Research: Solid Earth} {\bfseries 104} 29311}.

\bibitem{Lakes}
S.~Goeller, D.~Steinhage, M.~Thoma and K.~Grosfeld, \emph{Assessing the
  subglacial lake coverage of {A}ntarctica},
  \href{https://doi.org/10.1017/aog.2016.23}{\emph{Annals of Glaciology}
  {\bfseries 57} (2016) 109–117}.

\bibitem{MACGREGOR2009222}
J.~MacGregor, K.~Matsuoka and M.~Studinger, \emph{Radar detection of accreted
  ice over {L}ake {V}ostok, {A}ntarctica},
  \href{https://doi.org/https://doi.org/10.1016/j.epsl.2009.03.018}{\emph{Earth
  and Planetary Science Letters} {\bfseries 282} (2009) 222 }.

\bibitem{doi:10.1029/2005JD006349}
P.~R.~F. Barnes, E.~W. Wolff and R.~Mulvaney, \emph{A 44 kyr paleoroughness
  record of the {A}ntarctic surface},
  \href{https://doi.org/10.1029/2005JD006349}{\emph{Journal of Geophysical
  Research: Atmospheres} {\bfseries 111} }.

\end{thebibliography}\endgroup

\end{document}